# Optimal Stochastic Dynamic Scheduling for Managing Community Recovery from Natural Hazards

Saeed Nozhati[a,*], Yugandhar Sarkale[b], Edwin K. P. Chong[b,c], Bruce R. Ellingwood[a]

[a]Dept. of Civil and Environmental Engineering, Colorado State University, Fort Collins, CO 80523-1372
[b]Dept. of Electrical and Computer Engineering, Colorado State University, Fort Collins, CO 80523-1373
[c]Dept. of Mathematics, Colorado State University, Fort Collins, CO 80523-1874

## Abstract

Following the occurrence of an extreme natural or man-made event, community recovery management should aim at providing optimal restoration policies for a community over a planning horizon. Calculating such optimal restoration polices in the presence of uncertainty poses significant challenges for community leaders. Stochastic scheduling for several interdependent infrastructure systems is a difficult control problem with huge decision spaces. The Markov decision process (MDP)-based optimization approach proposed in this study incorporates different sources of uncertainties to compute the restoration policies. The computation of optimal scheduling schemes using our method employs the *rollout* algorithm, which provides an effective computational tool for optimization problems dealing with real-world large-scale networks and communities. We apply the proposed methodology to a realistic community recovery problem, where different decision-making objectives are considered. Our approach accommodates current restoration strategies employed in recovery management. Our computational results indicate that the restoration policies calculated using our techniques significantly outperform the current recovery strategies. Finally, we study the applicability of our method to address different risk attitudes of policymakers, which include risk-neutral and risk-averse attitudes in the community recovery management.

## 1    Introduction

Natural and man-made hazards pose significant challenges to civil infrastructure systems. Although proactive mitigation planning may lessen catastrophic effects, efficacious recovery scheduling can yield significant post-event benefits to restore functionality of critical systems to a level of normalcy in a timely fashion, thereby minimizing wastage of limited resources and disaster-related societal disorders. During the recovery process, the decision maker (also called "agent") must select recovery actions sequentially to optimize the objectives of the community. There are several characteristics of a *rational* agent and selecting a decision-making approach can become complicated. The most important characteristics of a rational decision-making approach include:

i.   The agent must balance the reluctance for low immediate reward with the desire of high future rewards (also referred as "non-myopic agent" or look-ahead property);
ii.  The agent must consider different sources of uncertainties;
iii. The agent must make decisions periodically to not only take advantage of information that becomes available when recovery actions are in progress but also to adapt to disturbances over the recovery process;
iv.  The agent must be able to handle a large decision-making space, which is typical for the problems at the community level. This decision-making space can cause an agent to suffer from *decision fatigue*; no matter how rational and high-minded an agent tries to be, one cannot make decision after decision without paying a cost [1].
v.   The agent must consider different types of dependencies and interdependencies among networks, because a single decision can trigger cascading effects in multiple networks at the community level.
vi.  The agent must be able to handle multi-objective tasks, which are common in real-world domains. The interconnectedness among networks and probable conflicts among competing objectives complicate the decision-making procedure.
vii. The agent must consider different constraints, such as time constraints, limited budget and repair crew, and current regional entities' policies.
viii. External factors, like the available resources and the type of community and hazard, shape the risk attitude of the agent. The different risk behaviors must be considered.

Community-level decision makers would benefit from an algorithmic framework that empowers them to take rational decisions and that accounts for the characteristics above. Markov Decision Processes (MDPs) address stochastic dynamic decision-making problems efficiently and offer an agent the means to identify optimal sequential post-event restoration policies.

In the realm of civil infrastructure management, several studies have used MDPs to optimize the repair and maintenance of infrastructure [2-4]. Papakonstantinou and Shinozuka [5] reviewed the literature on optimal maintenance planning using Dynamic Programming (DP) and MDPs. Madanat [6] introduced Latent MDP (LMDP) at the component level to recognize the random errors in the measurements of the conditions of infrastructure. Subsequently, Smilowitz and Madanat [7] extended LMDP to system-level maintenance scheduling, where they considered condition state and budgetary constraints. Medury and Madanat [8] used Approximate Dynamic Programming (ADP) with MDP for pavement management systems. Meidani and Ghanem [9] studied the problem of maintenance of pavement using DP and MDP with random transitions.

This study introduces a stochastic scheduling formulation based on MDP to identify near-optimal recovery actions following extreme natural hazards. This approach can support rational risk-informed decision making at the community level. To this end, we leverage the ADP paradigm to address large-scale scheduling problems in a way that overcomes the notorious *curse of dimensionality* that challenges the practical use of Bellman's equation [11]. We employ a promising class of approximation techniques called *rollout* algorithms. The application of ADP and rollout algorithms, along with the MDP formulation, provides not only a robust and computationally tractable approach but also the flexibility of incorporating current organizational recovery policies. In addition, we show how to treat current restoration policies as heuristics in the rollout mechanism.

As an illustrative example, we consider critical infrastructure systems within a community modeled after Gilroy, California, which is susceptible to severe earthquakes emanating from the San Andreas Fault. We model the Electrical Power Network (EPN), Water Network (WN), and main food retailers, including interconnectedness within and between networks. The EPN is particularly critical because the restoration and operation of most other vital systems need electricity. Additionally, the WN and food retailers supply water, food (e.g., ready-to-eat meals), and prescription medications that are essential for human survival following disasters. The functionality of the WN not only depends on its physical performance but also on the operation of the EPN, where a working EPN provides electricity for pumping station and water tanks. The serviceability of food retailers depends heavily on the WN and EPN. We consider these interdependencies and define two decision-making objective functions for optimization: to minimize the number of days needed to restore networks to an arbitrary level of service and to maximize number of people who have utilities per unit of time. We show how the proposed approach enables the agent (decision maker) to compute near-optimal recovery strategies to provide the three essential services — electricity, potable water, and food — to urban inhabitants and food retailers following a severe earthquake. We discuss the integrated recovery policies that consider multiple networks and objectives simultaneously, which can remarkably outperform the conventional isolated policies. Finally, we also discuss how risk-averse decision makers can utilize the proposed method.

## 2 Technical Preliminaries

In this section, we present the mathematical setting for the MDP. A detailed treatment of the subject is available in [10].

### 2.1 MDP Framework

A Markov decision process (MDP) is defined by the six-tuple $(X, A, A(.), P, R, \gamma)$, where $X$ denotes the state space, $A$ denotes the action space, $A(x) \subseteq A$ is the set of admissible actions in state $x$, $P(y \mid x, a)$ is the probability of transitioning from state $x \in X$ to state $y \in X$ when action $a \in A(x)$ is taken, $R(x, a)$ is the reward obtained when action $a \in A(x)$ is taken in state $x \in X$, and $\gamma$ is the discount factor. Let $\Pi$ be the set of Markovian policies ($\pi$), where $\pi: X \rightarrow A$ is a function such that $\pi(x) \in A(x)$ for each $x \in X$. Our goal is to compute a policy $\pi$ that optimizes the *expected total discounted reward* given by

$$V^{\pi}(x) := E\left[\sum_{t=0}^{\infty} \gamma^t R(x_t, \pi(x_t)) \,\middle|\, x_0 = x\right]. \tag{1}$$

The *optimal value function* for a given state $x \in X$ is connoted as $V^{\pi^*} : X \rightarrow \Re$ given by

$$V^{\pi^*}(x) = \sup_{\pi \in \Pi} V^{\pi}(x). \tag{2}$$

The optimal policy is given by

$$\pi^* = \arg\sup_{\pi \in \Pi} V^{\pi}(x). \tag{3}$$

Note that the optimal policy is independent of the initial state $x_0$. Also, note that we maximize over policies $\pi$, where at each time $t$ the action taken is $a_t = \pi(x_t)$.

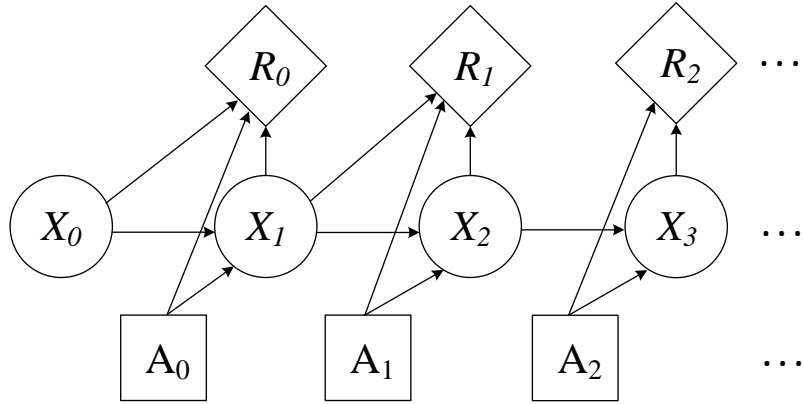

**Figure 1. Decision graph of a MDP**

The optimal policy $\pi^*$ can be computed using different methods, which include linear programming and dynamic programing. The methods of value iteration, policy iteration, policy search, etc., can find a strict optimal policy. We briefly discuss the Bellman's optimality principle [11], useful for defining the $Q$-value function, which plays a pivotal role in the description of the rollout algorithm. The Bellman's optimality principle states that $V^{\pi^*}(x)$ satisfies

$$V^{\pi^*}(x) := \sup_{a \in A(x)} \left\{ R(x,a) + \gamma \sum_{y \in X} P(y \mid x,a) V^{\pi^*}(y) \right\}. \tag{4}$$

The Q-value function associated with the optimal policy $\pi^*$ is defined as

$$Q^{\pi^*}(x,a) := R(x,a) + \gamma \sum_{y \in X} P(y \mid x,a) V^{\pi^*}(y), \tag{5}$$

which is the term within the curly braces in (4). Similarly, we can define $Q^{\pi}(x,a)$ associated with any policy $\pi$.

## 2.2 Simulation-based MDP

For large-scale problems, it is essential to represent the state, action, or outcome space in the framework defined in Section 2.1 in a compact form. Such a compact representation is possible in the simulation-based representation [12]. A simulation-based representation of an MDP is a 7-tuple $(X, A, A(.), P, R, \gamma, I)$, where $|X|$ or $|A|$ ($|\cdot|$ = the cardinality of the argument set "$\cdot$") is usually large, and the matrix representation of $P$ and $R$ is infeasible because of the large dimensions of a typical community recovery problem. We represent $P$, $R$, and $I$ as functions implemented in an arbitrary programming language. $R$ returns a real-valued reward, given the current state, current action, and future state ($R: X \times A \times X \to \Re$). $I$ is a stochastic function that provides state according to the initial state distribution. $P$ is a function that returns the new state given the current state and action. Essentially, the underlying MDP model is implemented as a simulator.

## 2.3 Approximate Dynamic Programming

Calculating an optimal policy using the methods above is usually infeasible owing to the dimensions of the state and/or action spaces. The size of the state/action space grows exponentially with the number of *state/action variables*, a phenomenon referred to by Bellman as the *curse of dimensionality*. The computational costs of running a single iteration of the value iteration and policy iteration algorithm, for the MDPs defined in Section 2.1, are $O(|X|^2|A|)$ and $O(|X|^2|A| + |X|^3)$, respectively. The computational cost of finding the optimal policy by directly solving the linear system provided by the Bellman equation is $O(|X|^3 |A|^3)$. Additionally, the computational cost of an exhaustive direct policy search algorithm, for a single trajectory consisting of $K$ simulation steps, is $3\left(\sum_{k=1}^{K} |X|^k\right)|A|^{|X|}|X|$ [13], which is prohibitive for even small-sized problems.

These computationally intractable algorithms cannot be used for large problems involving resilience assessment or recovery of a real-size community and approximate solutions are necessary. To this end, several algorithms have been developed in the realm of Approximate Dynamic Programming (ADP) that result in tractable computations for finding the near-optimal restoration policies. One popular class of algorithms involves approximating the $Q$-value function in (5). However, it often is difficult in practice to identify a suitable approximation to the $Q$-value function for practical, large-scale problems. In the following, we pursue a promising class of ADP algorithms known as *rollout* that sidesteps these difficulties by avoiding an explicit representation of the $Q$-value function.

## 2.4 Rollout

While computing an optimal policy for an MDP is often quite difficult because of the curse of dimensionality, policies based on heuristics (termed as *base policies*) can be readily designed in many cases. The principal idea behind the rollout technique is to improve upon the performance of the base policy through various means. Therefore, the base policy does not have to be close to optimal. In this study, we focus on improvement of the base policy through simulation. The idea was first proposed for stochastic scheduling problems by Bertsekas and Castanon [14]. Instead of the classical DP scheme [15], the agent "rolls out" or simulates the available policy over a selected finite horizon $H < \infty$; thereafter, the agent implements the most "promising" action in an *on-line* fashion, which determines optimal actions only for states that can be realized in the real world (reachable states), thereby conserving computational effort on unreachable states.

Monte Carlo (MC) simulations assess the $Q$-value function on demand. To estimate the $Q$-value function ($\hat{Q}^\pi(x, a)$ represents the estimate) of a given state-action pair $(x, a)$, we simulate $N_{MC}$ number of trajectories, where each trajectory is generated using the policy $\pi$, has length $H$, and starts from the pair $(x, a)$. The assessed $Q$-value function is typically taken as the average of the sample returns obtained along these trajectories:

$$\hat{Q}^\pi(x, a) = \frac{1}{N_{MC}} \sum_{i_0=1}^{N_{MC}} \left[ R(x, a, x_{i_0,1}) + \sum_{k=1}^{H} \gamma^k R(x_{i_0,k}, \pi(x_{i_0,k}), x_{i_0,k+1}) \right]. \tag{6}$$

For each trajectory $i_0$, we fix the first state-action pair to $(x, a)$; the simulator provides the next state $x_{i_0,1}$ when the current action $a$ in state $x$ is completed. Thereafter, we choose actions using the base policy. Note that if the simulator is deterministic, a single trajectory suffices, while in the stochastic case, a sufficient number of trajectories ($N_{MC}$) should be pursued to approximate the $Q$-value function. In this study, we focus on the rollout policy computed with single-step look-ahead. An agent can consider multistep look-ahead, at an added computational cost, to extract maximum performance out of our solution technique. The number of look-ahead steps mainly depends on the scale of the problem, computational budget, real-time constraints, and agent's preferences.

An important property of the rollout algorithm is that it improves upon the performance of the underlying base policy, if the base policy is not *strictly* optimal. The rollout policy computed using our method is not necessarily strict-optimal, but it is guaranteed that it would never perform worse than the underlying base policy. In fact, our simulation results suggest huge improvements over the base policy. Our framework offers the agent the flexibility of incorporating the current regional entities' policy as the base policy. To properly tailor the MDP and rollout formulations for our problem, we describe the real test-bed community and infrastructure systems in the next section and the proposed formulation in Section 4.

## 3    Community Testbed

To evaluate the applicability and efficiency of the proposed methodology on real communities, we model the community in Gilroy, California, USA, which is susceptible to severe earthquakes on the San Andreas Fault. The City of Gilroy is a moderately-sized growing city and located approximately 50 kilometers (km) south of the city of San Jose. It had a population of 48,821 at the time of the 2010 census. The availability of reasonable information about EPN, WN, population density, main food retailers, as well as the high level of seismic exposure of the area makes it an interesting case study. The study area is divided into 36 rectangular regions organized as a grid to define the properties of community, with an area of 41.9 km$^2$ and with a population of 47,905 as shown in Fig. 2. In this section, we provide the general information about the studied networks and facilities of the community. Detailed information of the community is provided in [16, 17].

## 3.1 Electrical Power and Water Networks

Figs. 3 and 4 show, respectively, the EPN and WN of Gilroy community. A 115 kV transmission line supplies the Llagas power substation, which provides electricity to the distribution system. The distribution line components are placed at intervals of 100 m and modeled from the power substation to the main facilities of WN, food retailers, and the centers of urban grids. The EPN and WN in this study are described as networks that contain nodes (vertices) and links (edges). The dependencies within and between networks are modeled through an adjacency matrix $\mathbf{A} = [x_{ij}]$, where $x_{ij} \in [0, 1]$ indicates the magnitude of dependency between components $i$ and $j$ [18]. If a component is damaged following a hazard, it becomes non-functional and all dependent components in the same or other networks would also be non-functional. The adjacency matrix $\mathbf{A}$ captures these cascading effects.

The EPN does not depend on any other network; hence, we only need to consider the dependency within the network. The probability that a critical facility like a water pump or a food retailer $G$ has electricity is

$$P(EG) := P\left( \bigcap_{l=1}^{\hat{n}} EE_l \right), \tag{7}$$

where $EG$ is the event that $G$ has electricity, $EE_l$ is the event that the $l^{th}$ EPN component is functional, and $\hat{n}$ is the minimum number of EPN components required to supply electricity to $G$. The sample space is a singleton set that has the outcome, "is functional." The $l^{th}$ EPN component is functional when it is undamaged or completely repaired, *and* all the EPN components serving that EPN component are functional. Fig. 3 shows the interdependence between the EPN components.

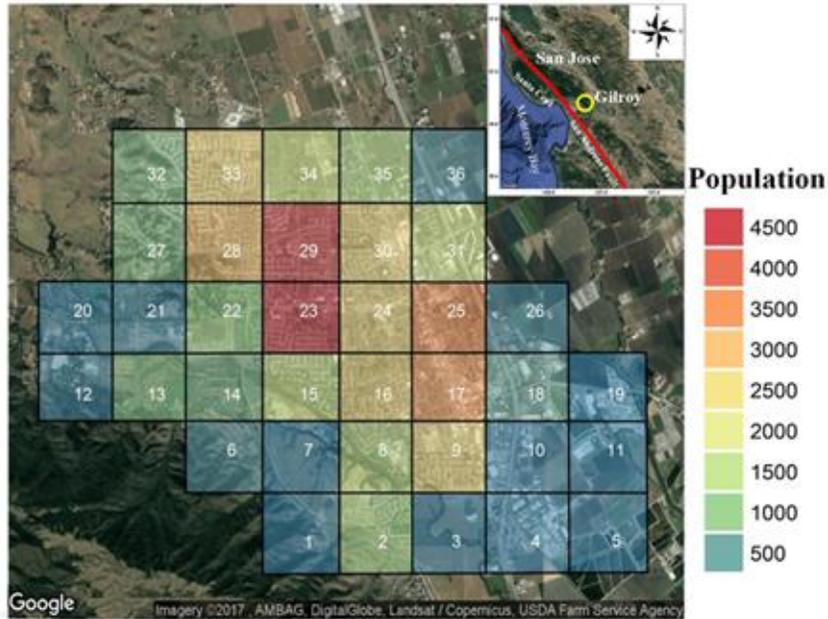

**Figure 2: Map of Gilroy's population over the defined grids**

The municipal water network of Gilroy is supplied by the Llgas sub-basin, which is recharged by Llagas and Uvas Creeks [16]. We have considered the main potable water pipelines, wells, water tanks, and booster pump stations (BPS) of the WN, as shown in Fig. 4. The functionality of the WN depends on not only the functionality of its components but also the availability of electricity following the extreme hazard for water pump operation. Therefore, we consider the dependency between WN and EPN. The probability that a critical facility like an urban grid rectangle or a food retailer $G$ has potable water is

$$P(WG) := P\left(\bigcap_{l'=1}^{\hat{k}} EW_{l'}\right), \tag{8}$$

where $WG$ is the event that $G$ has water, $\hat{k}$ is the minimum number of WN components (which could be BPS, wells, water tanks, or pipelines) required to supply water to $G$, and $EW_i$ is the event that the $l'^{th}$ WN component is functional. Again, the sample space here is a singleton set that has the outcome, "is functional." If the $l'^{th}$ WN component is a pipeline, then it is said to be functional when it is undamaged or completely repaired, *and* all the WN components serving that pipeline are functional.

In addition, if the $l'^{th}$ WN component is a BPS, well, or water tank, then it is said to be functional when it is undamaged or completely repaired *and* all the EPN components serving the $l'^{th}$ WN component are functional. While the pipelines are not directly dependent on the EPN, they are indirectly dependent on the EPN through the other WN components. The variables $\hat{n}$ and $\hat{k}$ accommodate any potential redundancy in the EPN and WN.

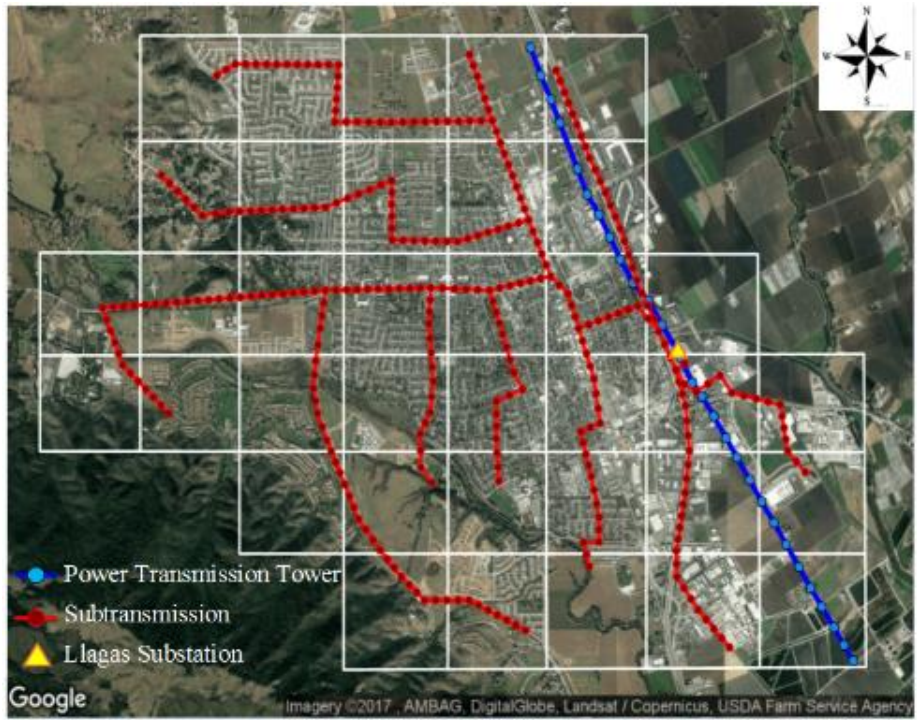

**Figure 3: Gilroy electrical power network**

## 3.2. Food Retailers

Food retailers play a vital role in the well-being of households in Gilroy. Fig. 5 shows the locations of the six main food retailers of Gilroy (each of which has more than 100 employees) that provide services to the community. Electricity and water are two essential elements that play crucial role in the functionality of retailers following hazards. Policymakers are always concerned about the adequate supply of these critical utilities for retailers. Hence, we consider food retailers and their dependencies to EPN and WN during restoration analysis and optimization. To capture the effects of the restoration of each food retailer on different households over the community, we use a gravity model [19, 20], which assigns the shopping probabilities based on the food retailers' capacities and locations so that bigger and closer retailers have greater impacts on household units.

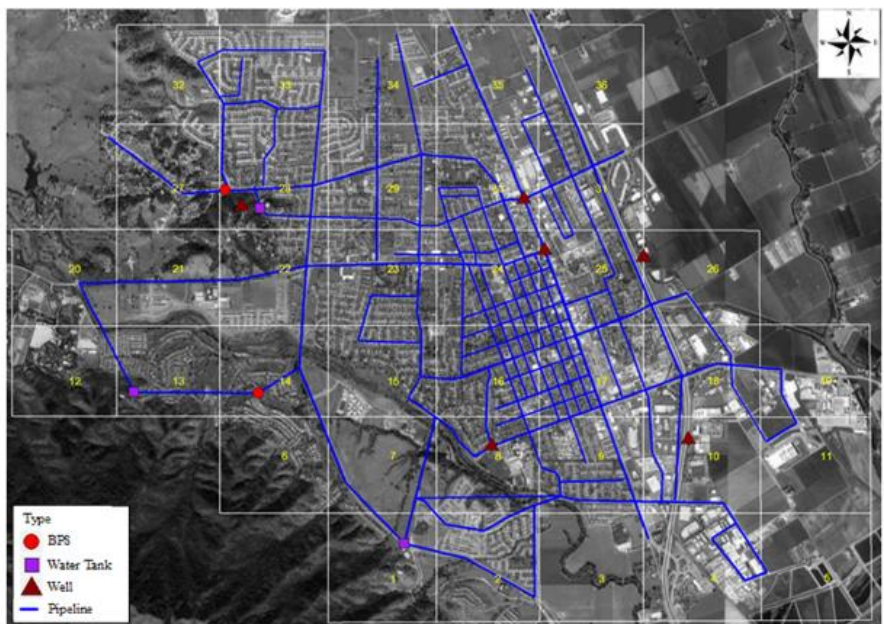

**Figure 4: Gilroy water network**

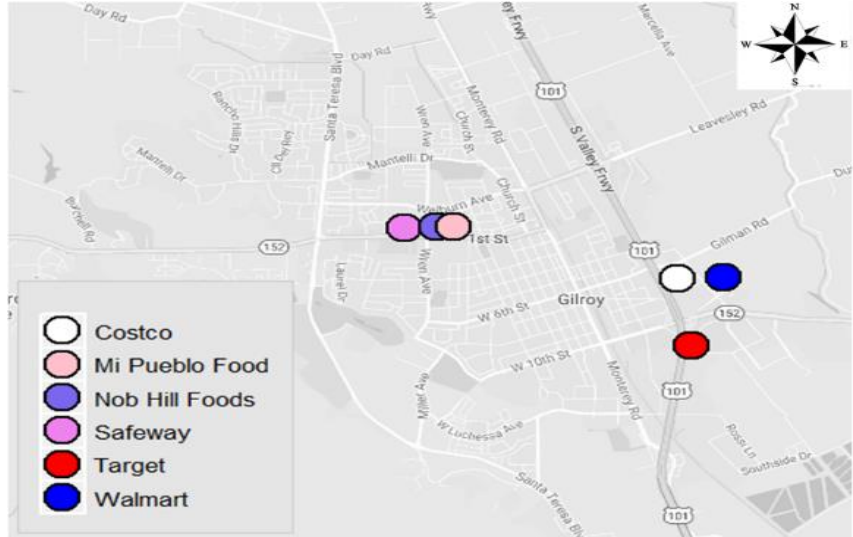

**Figure 5: Gilroy's main food retailers**

## 3.3. Hazard Modeling

Gilroy is susceptible to the effects of severe earthquakes from the nearby San Andreas Fault (SAF). The event simulated for this study is an earthquake of magnitude $M_w$=6.9 that occurs at one of the closest points on the SAF projection to downtown Gilroy with an epicentral distance of roughly 12 km. We use the Abrahamson et al. [21] Ground Motion Prediction Equation (GMPE) to estimate the median seismic demands (Intensity Measures) on infrastructure facilities: Peak Ground Acceleration (PGA) for the EPN components and above-ground WN facilities and Peak Ground Velocity (PGV) for buried pipelines. We assess the physical damage to components with seismic fragility curves [22-25].

Available repair crews, replacement components, and required tools for restoration are designated as Resources Units (RU). The number of RUs depends on the capacities and policy of the regional entities following disasters. Restoration times are synthesized based on exponential distributions from studies [20, 22-25], as summarized in Table 1. The proposed framework, nevertheless, allows one to use any arbitrary distribution.

**Table 1: The expected repair times (Unit: days)**

| | Damage states | | | |
|---|---|---|---|---|
| Component | Minor | Moderate | Extensive | Complete |
| Electric Sub-station | 1 | 3 | 7 | 30 |
| Transmission line component | 0.5 | 1 | 1 | 2 |
| Distribution line component | 0.5 | 1 | 1 | 1 |
| Water tanks | 1.2 | 3.1 | 93 | 155 |
| Wells | 0.8 | 1.5 | 10.5 | 26 |
| Pumping plants | 0.9 | 3.1 | 13.5 | 35 |

# 4    Post-hazard Recovery Formulation

Following an earthquake, the EPN and WN systems and components either remain undamaged or exhibit a level of damage, which is determined from the seismic fragility curves. Suppose that an agent must restore a community that includes several networks, which function as a System of Systems (SoS). Let $L^{'}$ be the total number of damaged components at time $t$, and let $t_c$ denote the decision time at which all the damaged components are repaired ($L^{'}=0$). The agent has only a limited number of RUs that can be assigned, usually much less than $L^{'}$, especially in severe disasters that impact large communities. The RUs differ from network to network because of the skill of repair crews and qualities of the required tools. The problem is to assign the available RUs to $L^{'}$ damaged components in a manner that best achieves the community objectives and policymakers' preferences.

We make the following assumptions: (1) The agent has access to all the damaged component for repair purposes; (2) A damaged component only needs one RU to be repaired and assigning more than one RU would not reduce the repair time [26]; (3) The agent has limited RUs for each network and cannot assign a RU of one network to another (e.g., a WN RU cannot be assigned to the EPN); (4) The agent can preempt the assigned RUs from completing their work and reassign them at different locations to maximize the beneficial outcomes. (5) Once a damaged component is repaired, all assigned RUs are available for re-assignment even if their assigned components are not fully repaired. It is also possible to let the RU continue the repair work at the same location in the next time slot according to the objectives of the agent. We refer to such assignment as *preemptive scheduling*, which allows the agent to be flexible in planning and is particularly useful when a central stakeholder manages an infrastructure system; see [38] for a discussion on *non-preemptive scheduling*. (6) The agent can deal with stochastic scheduling, where the outcome of the repair actions is not fully predictable and can be quantified probabilistically. The unpredictability mainly arises from the randomness in the repair times (see Table 1). The MDP simulator exhibits stochastic behavior owing to the random repair times. On the other hand, the alternative perspective, where the outcome of actions is fully predictable, is also an active research topic [20, 36, 37].

## Markov Decision Process Formulation

Suppose that $x_t^E$ and $x_t^W$, respectively, represent the damage state of the EPN and WN at time $t$. $x_t^E$ is a vector of length $L_t^E$, where $L_t^E$ is the number of damaged components in the EPN. Each element of the vector $x_t^E$ is in one of the five damaged states (counting no damage as one state) in Table 1. Similarly, we define $x_t^W$ of length $L_t^W$, where $L_t^W$ is the number of damaged components in the WN at time slot $t$. Let $N_E$ and $N_w$ denote the available RUs for the EPN and WN, respectively, with $N_E \leq L_1^E$ and $N_W \leq L_1^W$. We can define the tuples of our MDP framework as follows:

- *States X*: $x_t$ denotes the state of the damaged components in the community at time slot $t$ as the stack of two vectors, $x_t^E$ and $x_t^W$, as follows:

$$x_t := \left( x_t^E, x_t^W \right) \quad s.t. \quad |x_t| = L_t^E + L_t^W. \qquad (9)$$

- *Actions A*: $a_t$ denotes the repair actions to be carried out on the damaged components at time slot $t$, as the stack of two vectors, $a_t^E$ and $a_t^W$ ,

$$a_t := \left( a_t^E, a_t^W \right) \quad s.t. \quad |a_t| = L_t^E + L_t^W, \qquad (10)$$

where both $a_t^E$ and $a_t^W$ are binary vectors of length $L_t^E$ and $L_t^W$ , respectively, where a value of zero means no repair and one means carry out repair action. $a_t^E$ and $a_t^W$ represent the actions (no repair, repair) to be performed on the damaged components of the EPN and WN.

- *Set of Admissible Actions $A(x_t)$*: The set of admissible repair actions $A(x_t)$ for the state $x_t$ is the set of all possible binary combinations of integers one and zero such that each element of this set is of size $L_t^E + L_t^W$ , and each element has $N_E$ number of ones in the first $L_t^E$ locations and $N_w$ number of ones at the remaining locations. The interdependence between networks explodes the size of the set of admissible actions as follows:

  Let $D_t^E$ be the set of all damaged components of the EPN before a repair action $a_t$ is performed. $P\left( D_t^E \right)$ denotes the powerset of $D_t^E$ ;

  $$P_{N_E}\left( D_t^E \right) := \left\{ C \in P\left( D_t^E \right) : |C| = N_E \right\}, \qquad (11)$$

  where $\left| P_{N_E}\left( D_t^E \right) \right|$ represents the size of the set of admissible actions for the EPN. We can also define $P_{N_W}\left( D_t^W \right)$ similarly. The size of the set of admissible actions, at any time $t$, is the product of the size of set of admissible actions for EPN and WN:

  $$|A(x_t)| := \left| P_{N_E}\left( D_t^E \right) \right| \times \left| P_{N_W}\left( D_t^W \right) \right| . \qquad (12)$$

  Therefore, when multiple networks are considered simultaneously, the size of $A(x_t)$ grows very quickly. Searching exhaustively over the entire set $A(x_t)$ for calculating the optimal solution is not possible; therefore, we employ the rollout technique, as described in Section 2.4.

- *Simulator P*: Given, $x_t$ and $a_t$, the simulator $P$ provides the new state $x_{t+1}$. $P$ is a *generative model* that can be implemented as a simulator, without any explicit knowledge of the actual transitions. It considers the interconnectedness within and between networks to compute the cascading effects of $a_t$ through the whole community and recovery process. As we alluded to before, a compact representation of $P$ is important for large-scale problems.

  In our problem formulation, as soon as at least one of the damaged components is repaired, the repair action $a_t$ is considered complete. Define this completion time at every $t$ by $\hat{t}_t$. Recall that the repair time is exponentially distributed. The completion time is the minimum of the repair time at one or more damaged locations, where a repair action is being performed. The minimum of exponential random variables is exponentially distributed; therefore, the completion time is also exponentially distributed [27]. The sojourn time (a.k.a. the holding time) is the amount of time that the system spends in a specific state. For an MDP, the sojourn time, $t_s$, is exponentially distributed [15, 27, 28]. Note that for our MDP formulation, $\hat{t}_t$ is equal to $t_s$.

A natural question that arises is "does this formulation work when the repair times are non-exponential?" In that case, the completion time is not exponentially distributed. However, in our present problem formulation, the completion time is the same as the sojourn time. Thus, the sojourn time would not be exponentially distributed, which is inconsistent with the Markovian assumption. This can be remedied simply by incorporating the lifetime of the damaged component into the state definition. The lifetime of the damaged component is the time required for the damaged component to be repaired after the occurrence of hazard. With this new definition of the state space, the sojourn time is different from the completion time $\hat{t}_t$, and the sojourn time is exponentially distributed. Here the completion time $\hat{t}_t$ is still the minimum of the repair time at one or more damaged locations but with any underlying distribution of the repair times. Thus, our framework is sufficiently flexible to accommodate repair times with any underlying distribution.

- *Rewards R*: In this study, we pursue two different objectives for the agent.

  The first objective (hereinafter Obj. 1) is to optimally plan decisions so that a certain percentage of the total inhabitants (denoted by threshold $\alpha$) are *benefitted from the recovery of utilities* in the shortest period of time, implying that household units not only have electricity and water but also have access to a functional retailer that has electricity and water. Conversely, even if a household unit has electricity and water and has access to a retailer that has electricity but not water, the household unit does not benefit from the recovery actions. The mapping of people in the gridded rectangle to a food retailer is determined by the gravity model. We aim to optimally plan the repair actions to minimize the time it takes to achieve the benefit from utilities to $\alpha$ percent of people. The reward function for the first objective is defined as:

$$R_1(x_t, a_t, x_{t+1}) = \hat{t}_t, \tag{13}$$

  The second objective (hereinafter Obj. 2) is to optimally plan decisions so that maximum number of inhabitants are *benefited from recovery of utilities* per unit of time (days, in our case). Therefore, in the second case, there are two objectives embedded in our reward as follows:

$$R_2(x_t, a_t, x_{t+1}) = \frac{r}{t_{rep}}, \tag{14}$$

  where $r$ is the number of people deriving benefit from utilities after the completion of $a_t$, and $t_{rep}$ is the total repair time to reach $x_{t+1}$ from any initial state $x_0$ (i.e., $t_{rep} = \sum \hat{t}_t$). Note that the reward function is stochastic because the outcome of the repair action is stochastic.

- Initial State *I*: As mentioned in Section 3, the stochastic damage model of the EPN and WN components can be obtained by the fragility curves. The initial damage states associated with the components will be provided by these models.

- *Discount factor $\gamma$*: In this study, we set the discount factor to be 0.99 [29]. This is a measure of how "far-sighted" the agent is in considering its decisions. The discount factor weighs the future stochastic rewards at each discrete time $t$.

# 5 Results and Discussion

We divide the presentation of our simulation results into two sections. The first section caters to risk-neutral decision makers, and the second section caters to risk-averse decision makers [32, 33]. Each of these sections is further divided into two sub-sections to demonstrate the performance of our method on two separate objectives functions. When *Objective 1* is considered, the reward function in our MDP is given by (13), while for *Objective 2*, the reward function of our MDP is given by (14). For all the simulation results presented henceforth, we selected $N_{MC}$ in (6) and (15) so that the standard deviation of the estimated $Q$-value $\hat{Q}^{\pi}(x, a)$ is below 0.05.

As mentioned in Section 2.4, the most feasible base policy for community recovery planning often is the current recovery strategy of regional responsible companies or organizations. However, there is no restriction on the selection of a policy as a base policy. We proposed the alternatives for the definition of base policies for recovery management problems in [20]. In this study, the base policy is defined based on expert judgment and importance analyses that prioritize the importance of components owing to their contribution to the overall risk. Specifically, the restoration sequence defined by our base policy for EPN is transmission line, power substation, and distribution lines to downtown and water pumps; similarly, the base policy for WN involves water wells, water tanks, BPS, and pipelines to downtown and food retailers.

## 5.1 Mean-based Stochastic Optimization

The mean-based optimization is suited to risk-neutral decision makers [9]. In this approach, the optimal policy is determined based on the optimization of the $Q$-value function, where the estimate of the $Q$-value function $\hat{Q}^{\pi}(x, a)$ is based on the mean of $N_{MC}$ trajectories, as demonstrated in (6). Calculating the $Q$-value based on the expected $Q$-value of $N_{MC}$ trajectories may not always be appropriate, especially in the case of risk-averse decision makers. However, it has been shown that the mean-based stochastic optimization approach can be appropriate when the objective function properly encodes the risk preferences of policymakers [15]. Nevertheless, we demonstrate the performance of our method when the decision maker has risk-averse attitude to planning in Section 5.2.

### 5.1.1 Implementation of Rollout Algorithm for Objective 1

The rollout algorithm with respect to Obj. 1 identifies recovery strategies to minimize the time it takes to provide the utilities to $\alpha$ percent of people in the community. In this formulation, the selection of $\alpha$ depends on the preferences of policymakers. For our simulation, we selected $\alpha=0.8$, implying that we want to provide the benefit of the utility recoveries to 80% of people in minimum amount of time.

Figure 6 shows the performance of the rollout and base polices for Objective 1. The rollout algorithm optimizes the restoration of two networks, EPN and WN, simultaneously to provide utilities for 80% of people in 19.3 days following the earthquake, while the base policy completes this task in 26.1 days. This 35% improvement over the entire recovery period signifies the performance of rollout at the community level. Figure 6 also highlights the look-ahead property of rollout. Although the base policy showed a better performance during the first 15 days following the earthquake, the rollout algorithm outperformed the base policy in the whole recovery. By selecting conservative repair decisions initially, rollout can balance the desire for low present cost with the undesirability of high future costs.

The performance of rollout on the individual food retailers is summarized in Table 2. Note that the base policy restored EPN and WN to Safeway, Nob Hill Foods, and Mi Pueblo Food faster than the rollout policy; however, the base policy is incapable of determining the recovery actions to balance the rewards so that 80% of people benefit from restoration of utilities (our true objective).

**Table 2: Performance of rollout vs. base policy for the first objective function for the individual retailers**

| Policy | Recovery time | Costco | Walmart | Target | Safeway | Nob Hill Foods | Mi Pueblo Food |
|--------|---------------|--------|---------|--------|---------|----------------|----------------|
| Base | 26.06 | 0.31 | 0.31 | 21.02 | 5.91 | 5.91 | 2.76 |
| Rollout | 19.23 | 0.31 | 0.31 | 15.95 | 18.33 | 18.33 | 8.01 |

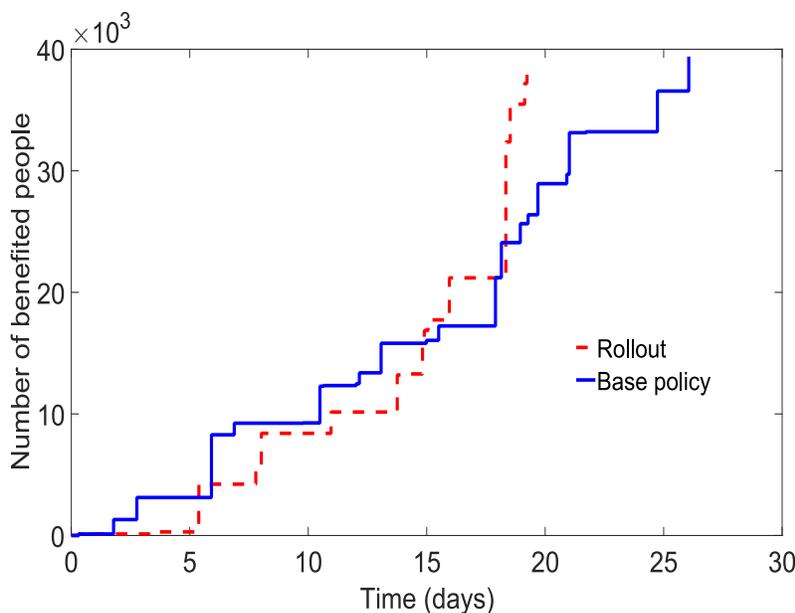

**Figure 6: Performance of rollout vs. base policy for the first objective function**

After 80% of the people have benefitted from utility restoration, we continue to evaluate the progress in restoration of the EPN and WN. Even though we have met our objective of providing the benefit of utilities to 80% of the population, 25% of the EPN components remain unrepaired. This interesting result shows the importance of prioritizing the repair of the components of the network so that the objectives of the decision maker are met. Because the objective here was to restore utilities so that 80% of people would benefit owing to the restoration in minimum amount of time, our algorithm prioritized repair of only those components that would have maximum effect on our objective without wasting resources on the repair of the remaining 25% of EPN components.

*5.1.2 Implementation of Rollout Algorithm for Objective 2*

The rollout algorithm applied to Objective 2 identifies recovery strategies that maximize the number of inhabitants per day that benefit from the strategy selected. In other words, the algorithm must maximize the area under the restoration curve normalized by the total recovery time. This objective function is specifically defined to match the definition of the common resilience index, which is proportional to the area under the restoration curve [30]. Fig. 7 depicts the performance of base policy and the corresponding rollout policy. The mean number of people that benefit from utility restoration based on the base policy is 22,395 per day, whereas that for the rollout policy is 24,224. These values are calculated by dividing the area under the curves in Fig. 7 by the total number of days for the recovery, which is our Obj. 2. Analogous to Fig. 6, Fig. 7 highlights the look-ahead property of the rollout algorithm for Obj. 2.

As in Section 5.1.1, we analyzed the performance of the rollout algorithm for the individual networks. One of the main reasons for this analysis is that these networks are restored and maintained by different public or private entities that would like to know how rollout would perform for their individual systems. We use the recovery actions $a_t$, computed using the rollout policy for the combined network that considers all the interdependencies (for Obj. 2), and check the performance of these repair actions on individual networks.

First, we check the performance of the repair actions on the EPN network, calculating the effect of EPN restoration on only the household units. The results are depicted in Fig. 8. The base policy leads to EPN recovery so that the mean number of people with electricity is 24,229 per day, while the rollout policy provides the electricity for 27,689 people on average. Second, we check the performance of the repair actions on the EPN, but considering the effect of EPN restoration on both household units and retailers. In this analysis, summarized in Fig. 9, people derive benefit of EPN recovery when their household unit has electricity and they go to a retailer that has electricity. In this case, the mean number of people who benefit from the EPN recovery owing to the base policy is 23,155/day, whereas that owing to the rollout policy is 25,906/day. Third, we check the performance of the repair actions on the WN, calculating the effect of WN restoration on only the household units, as illustrated in Fig. 10. In this case, the mean number of people with potable water under the base and rollout policies is 31,346/day and 25,688/day, respectively. Finally, we check the performance of the repair action on the WN, but where the effect of WN restoration on both household units and retailers is considered. In this case, people benefit from WN recovery when their household unit has water, and they go to a retailer that has water. In this case, the mean number of people with potable water under the base and rollout policies is 31,346/day and 25,688/day, as shown in Fig. 11.

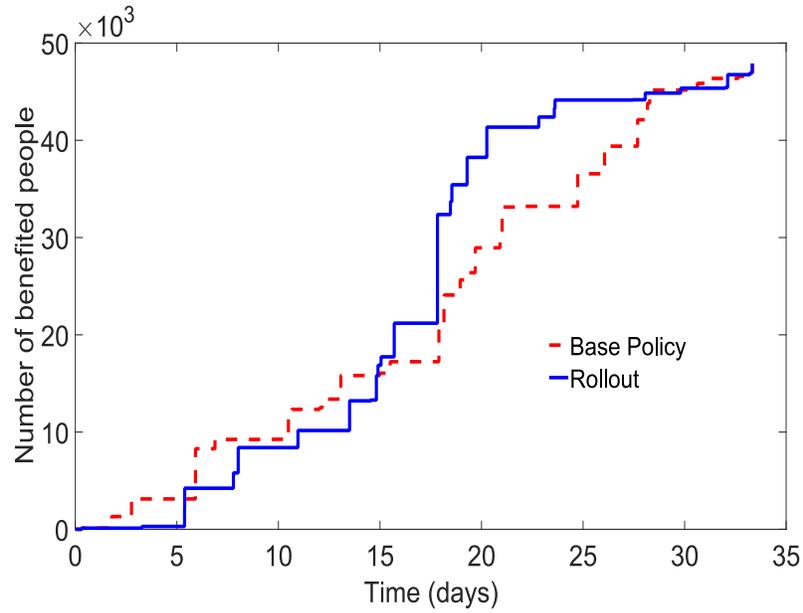

**Figure 7: Performance of rollout vs. base policy for the second objective function**

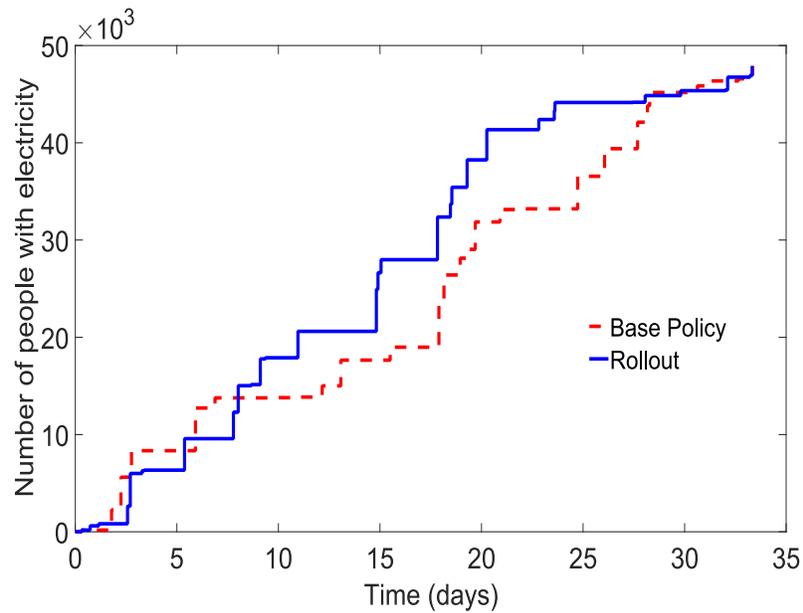

**Figure 8: The performance of policies to provide electricity for household units**

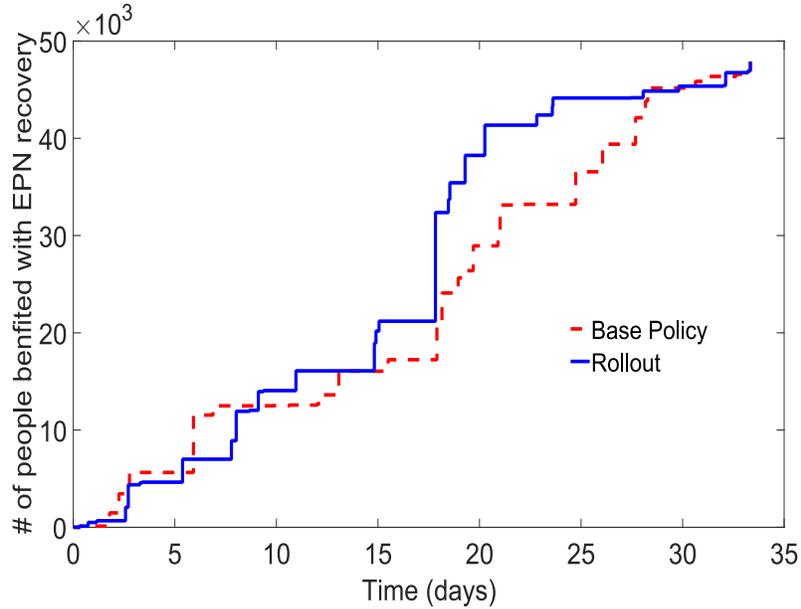

**Figure 9: The performance of policies to provide electricity for household units and retailers**

It is interesting to note that the rollout policy need not outperform the individual base policy when the recovery of each individual network is considered separately because in our framework, the calculation of recovery actions due to rollout considers the combined network and corresponding interdependencies that outperforms the base policy as shown in Figs. 6-9. Our objective considers two networks as one complex system (or SoS), which is captured in the definition of the benefit, and is not reflected in the restoration of a single network alone. Figs. 10 and 11 indicate that it is necessary to alleviate the concerns of individual stakeholders when recovery is performed based on interdependencies in the network. Refs. [20, 23] provide a thorough examination of the performance of rollout when the EPN and WN are considered separately. Furthermore, the number of days required to restore the WN is less than what is required to restore EPN, even when the optimized recovery actions for the combined network are used to evaluate the performance of the individual network restoration (see Figs. 8-11). This behavior can be attributed to a lesser number of WN components being restored compared to the number of EPN components.

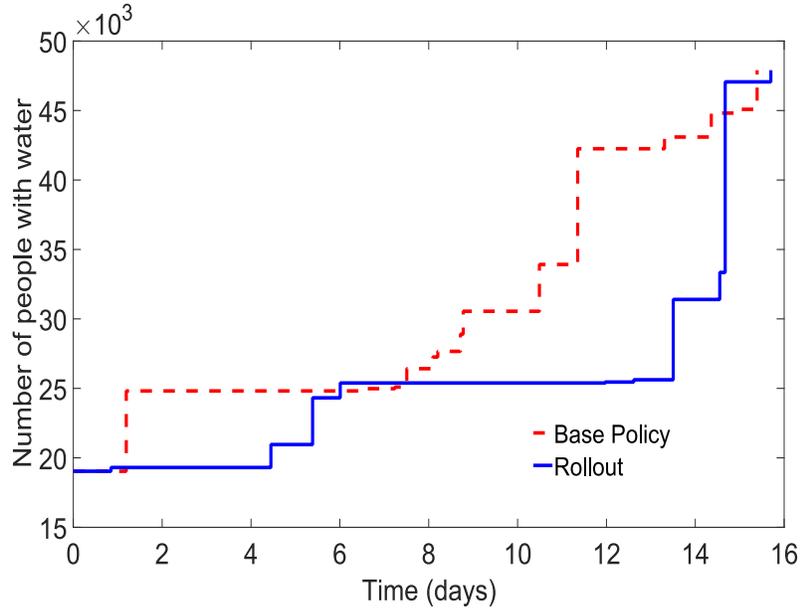

**Figure 10: The performance of policies to provide potable water for household units**

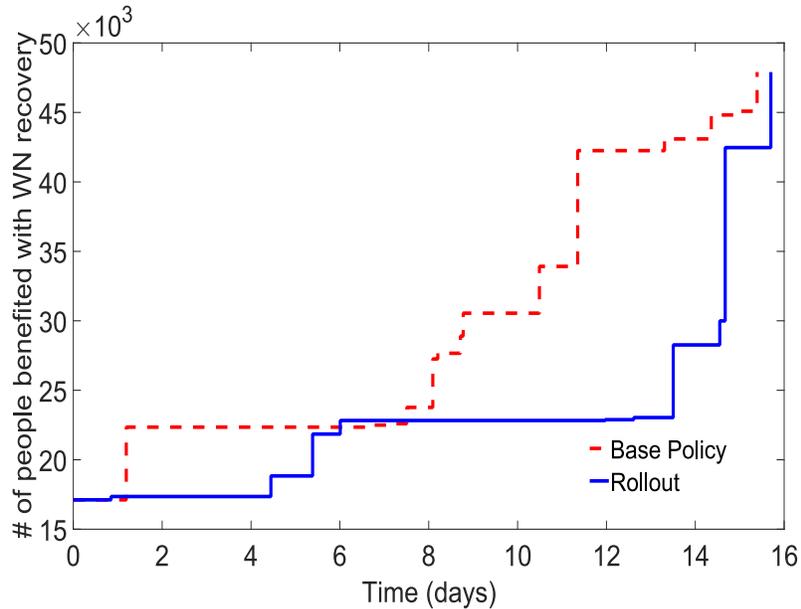

**Figure 11: The performance of policies to provide potable water for household units and retailers**

## 5.2 Worst-case Stochastic Optimization

The mean-based stochastic optimization seeks to identify the most cost-efficient repair actions in the face of uncertainty under the assumption that the decision maker has a risk-neutral attitude. This assumption has been criticized on several counts [31, 32]. Research on risk attitudes has revealed that most decision makers are not risk-neutral in the face of a low-probability threat or hazard. Moreover, policymakers and community stakeholders are not risk-neutral, especially when engaging large systems at the community level that influence public safety [33]. Finally, a stochastic model of uncertainty may not be possible in many practical problems in which only limited data exist and, accordingly, policy-makers tend to be more risk-averse [15]. These observations lead us to study the performance of the proposed rollout algorithm for risk-averse policymakers.

Risk-averse policymakers are more worried about extrema, rather than expected consequences of uncertainty. Worst-case optimization (a.k.a. robust optimization) is employed for MDPs to allow for risk-averse behavior [9, 34, 35]. Note that when Obj. 1 is under consideration, we are solving a minimization problem, whereas when Obj. 2 is under consideration, we deal with a maximization problem. As in Section 5.1, we make use of $N_{MC}$ trajectories. But unlike (6), we do not take mean of the $N_{MC}$ estimated $Q$-values to approximate the original $Q$-value function in (4) and (5). Instead, we use the maximum or minimum value among the $N_{MC}$ trajectories as a representation of worst-case behavior, depending on whether Obj. 1 or Obj. 2, respectively, is considered. If $i_0^*$ maximizes (6), where $i_0^* \in \{1, \ldots, N_{MC}\}$, then, for Obj. 1, the worst-case $Q$-value estimation is represented in (15). It is this estimated $Q$-value that is used in (4). Conversely, for Obj. 2, $i_0^*$ minimizes (6), where $i_0^* \in \{1, \ldots, N_{MC}\}$,

$$\hat{Q}^\pi(x,a) = R(x, a, x_{i_0^*,1}) + \sum_{k=1}^{H} \gamma^k R(x_{i_0^*,k}, \pi(x_{i_0^*,k}), x_{i_0^*,k+1}) \tag{15}$$

In the worst-case optimization simulations, when Obj. 1 is considered, the number of days required to reach the threshold of $\alpha = 0.8$ under the base policy is 26.1 days whereas under rollout, it is 19.7 days, a 32% improvement that signifies a desirable performance of the proposed methodology for the risk-averse policymakers. Fig. 12 shows the performance of rollout for Obj. 2, where the number of people deriving benefit from utilities per day because of recovery actions under the base and rollout policies is 22,395/day and 24,478/day. Fig. 12 also illustrates the look-ahead property, which is characteristic of the rollout algorithms. Finally, the performance of rollout for the individual networks is summarized in Table 3 and Fig. 13. As in Section 5.1.2, the results indicate that risk-averse policymakers should not presume that rollout will outperform the base policy when the EPN and WN are considered separately.

**Table 3: The performance of policies in different cases for the worse-case optimization (Unit: average No. of people per day)**

| Case | Base policy | Rollout policy |
|---|---|---|
| EPN restoration for household units | 24229 | 27897 |
| EPN restoration for household units and retailers | 23155 | 26159 |
| WN restoration for household units | 31346 | 25966 |
| WN restoration for household units and retailers | 30099 | 23535 |

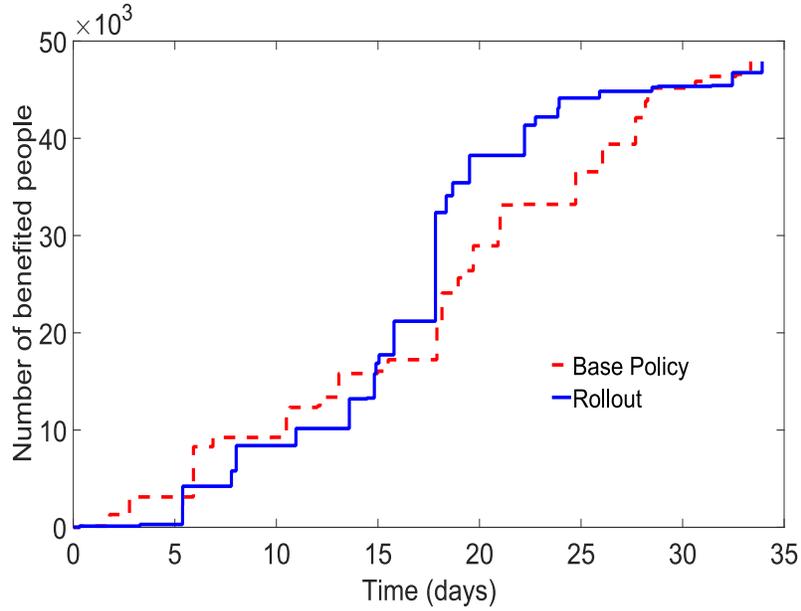

**Figure 12: Performance of rollout vs. base policy in the worst-case optimization for the second objective function**

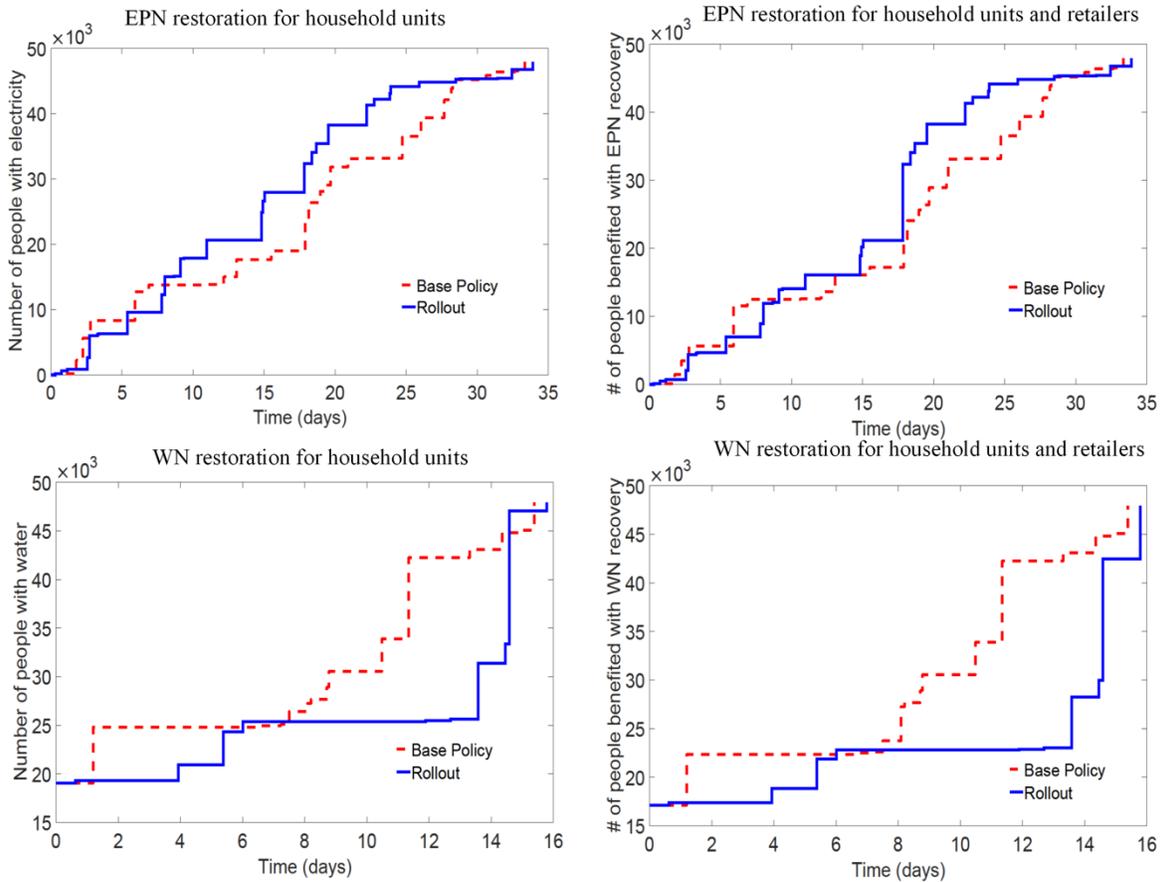

**Figure 13: The performance of policies in different cases for the worse-case optimization**

In summary, the results presented in this section advocate the desirable performance of the rollout algorithm in the face of a risk-averse attitude on the part of the decision maker. The individual attitudes toward risk can be dependent on the personalities of policymakers and stakeholders of a community and be influenced by many factors, such as the community properties, type of hazard, available resources and time, and existing information about the uncertainties to name a few. Lastly, because of the stochastic approximation involved in the computation of the estimated $Q$-values, it is not possible to compare the performance of the mean-based and worst-case optimization methods proposed above.

# 6    Conclusion

Community-level recovery was formulated as an MDP that accounts for different sources of uncertainties in the entire restoration process. Stochastic scheduling of community recovery that embeds several interconnected networks is a difficult stochastic control problem with huge decision spaces. As the computation of exact solutions for MDP is intractable for large problems, we utilized rollout algorithms, which fall under the broad umbrella of approximate dynamic programming techniques, for scheduling community-level recovery actions. The proposed methodology considers interdependent electrical and water networks in the community and treats them as one complex system. We tested the feasibility of the proposed method through a real case study involving a real community susceptible to severe earthquakes with respect to different objective functions that are popular for policymakers in the community resilience problems. We also considered the performance of the method for policymakers with different risk attitudes. The performance of the rollout policies appears to be near-optimal and is substantially better than the performance of their underlying base policies.

The proposed rollout approach has the all characteristics of a comprehensive framework, mentioned in Section 1. Furthermore, the rollout policy treats the community as a system of systems and provides the optimal strategies for the whole community. These strategies are not necessarily optimal for the individual networks and surely outperforms their underlying base policies.

## Acknowledgment

The research herein has been funded by the National Science Foundation under CRISP Collaborative Research Grant CMMI-1638284. This support is gratefully acknowledged. Any opinions, findings, conclusions, or recommendations presented in this material are solely those of the authors, and do not necessarily reflect the views of the National Science Foundation.